\documentclass[11pt,showpacs,amsmath,preprintnumbers,amssymb]{revtex4}
\usepackage[colorlinks=true,urlcolor=blue,linkcolor=blue]{hyperref}
\usepackage{graphicx}
 \relax

\usepackage{epsfig}
\usepackage{bm}

\newcommand{\rld}{\rho_{\Lambda D}}
\newcommand{\rl}{\rho_{\Lambda}}
\newcommand{\rlf}{\rho_{\Lambda 5}}
\newcommand{\oll}{\Omega_\Lambda}
\newcommand{\omm}{\Omega_M}

\begin{document}

\title{Holographic Dark Energy in Braneworld Models with a Gauss-Bonnet Term in the Bulk. Interacting Behavior and the $w =-1$ Crossing}

\author{E.~N.~Saridakis
\footnote{E-mail: msaridak@phys.uoa.gr}} \affiliation{Department
of Physics, University of Athens, GR-15771 Athens, Greece}

\begin{abstract}
We apply bulk holographic dark energy in general braneworld models
with a Gauss-Bonnet term in the bulk and an induced gravity term
and a perfect fluid on the brane. Without making any additional
assumptions we extract the Friedmann equation on the physical
brane and we show that a $\rho$-$\rho_\Lambda$ coupling arises
naturally by the full 5D dynamics. The low-energy (late-time)
evolution reveals that the effective 4D holographic dark energy
behaves as ``quintom'', that is it crosses the phantom divide
$w=-1$ during the evolution. In particular, the Gauss-Bonnet
contribution decreases the present value of $w_\Lambda$, while it
increases the growing rate of $w_\Lambda(z)$ with $z$, in
comparison with the case where such a term is absent.
\end{abstract}

\pacs{95.36.+x, 98.80.-k, 04.50.-h} \maketitle

\section{Introduction}

Holographic dark energy
\cite{Li,hol1,Gong,Guberina,Setare,Setare11} is an interesting and
simple idea of explaining the observed Universe acceleration
\cite{observ}. It arises when the more fundamental holographic
principle \cite{Hooft,witten0} is applied in the cosmological
framework \cite{Susskind.rh,holcosm} (although there are some
objections on this approach \cite{Linde}). Holographic dark energy
reveals the dynamical nature of the vacuum energy by relating it
to cosmological volumes. The background on which it is based, is
the black hole thermodynamics \cite{BH,5Dradius} and the
connection between the UV cut-of of a quantum field theory, which
is related to vacuum energy, and a suitable large distance of the
theory \cite{Cohen}. This connection, which was also known from
AdS/CFT correspondence, proves to be necessary for the
applicability of quantum field theory in large distances. The
reason is that while the entropy of a system is proportional to
its volume the black hole entropy is proportional to its area.
Therefore, the total energy of a system should not exceed the mass
of a black hole of the same size, since in this case the system
would collapse to a black hole violating the second law of
thermodynamics. When this concept is applied to the Universe, the
corresponding vacuum energy is the holographic dark energy.

Until now, almost all works on the subject have been formulated in
the standard 4D framework. On the other hand, brane cosmology
\cite{Rubakov83,RS99b} exhibits many phenomenological successes
\cite{branereview}. In a recent work \cite{manos.restored} we
presented a generalized and restored holographic dark energy in
the braneworld context. The basic argument was that in such a
framework black holes will in general be D-dimensional
\cite{BH,5Dradius} and therefore holographic dark energy should be
considered in the bulk. Subsequently, it gives rise to an
effective 4D dark energy with ``inherited" holographic nature, and
this one is present in the (also arisen from the full dynamics)
Friedmann equation of the brane. In \cite{manos.restored} we
applied this bulk holographic dark energy in a general
single-brane model and we reproduced the results of conventional
4D calculations \cite{Li,hol1,Gong,Guberina,Setare,Setare11},
having in mind that the physical interpretation is different. In
\cite{manos.movingbranes} we applied it in a general two-brane
model with moving branes and we showed that ``quintom'' behavior
\cite{quintom0,quintom,quintoma} arises naturally for a large
parameter space area of a simple solution subclass, without the
inclusion of special fields or potential terms. In particular we
found that $w_\Lambda$ was larger than $-1$ in the past while its
present value is $w_{\Lambda_0}=-1.08$, and the phantom divide
$w_\Lambda=-1$ was crossed at $z_{p}\approx0.49$, a result in
remarkable agreement with observations
\cite{observHDE1,observHDEaaa}.

In this work we examine general single-brane models, including a
Gauss-Bonnet term in the bulk
\cite{Charmousis,sigkr.liseis,TetradisGB,GBbulk.w1,GBbulk,Zamarias}
(see also \cite{GB4D.w1,GB4D} for a Gauss-Bonnet term in
conventional 4D cosmology). Such a higher-curvature combination
corresponds to the leading order quantum correction to gravity, in
an effective action approach to string theory and in particular in
the case of the heterotic string \cite{sloan}, and its coupling is
related to the Regge slope parameter on string scale. Furthermore,
the Gauss-Bonnet combination is the only curvature squared form
which gives ghost-free self-interactions for the graviton (around
flat spacetime) \cite{zwiebach} and maintains its zero modes of
the perturbations localized on the brane \cite{Abdesselam}.
Fortunately, holographic description holds for braneworld
Gauss-Bonnet gravity, although the subject is not trivial since
there are some ambiguities in the case of non-flat branes away
from the bulk boundary \cite{GBholography}. Applying bulk
holographic dark energy in this framework, and without any
additional assumption, we acquire the interesting situation of an
interaction between the 4D dark energy and the matter density of
the brane. In this case, cosmological evolution and in particular
the dependence of the 4D dark energy on the brane scale factor,
acquires a correction in terms of the Gauss-Bonnet coupling. The
rest of the text is organized as follows: In section \ref{HDEBulk}
we present the holographic dark energy in the bulk and in section
\ref{HDEBrane} we apply it to a general single-brane model in 4+1
dimensions with a Gauss-Bonnet term in the bulk. Finally, in
\ref{discussion} we discuss the physical implications of our
analysis and we summarize the obtained results.

\section{Formulation of Holographic Dark Energy in a General Bulk}\label{HDEBulk}

In this section we display the basic results of bulk holographic
dark energy, formulated in \cite{manos.restored}. The mass
$M_{BH}$ of a spherical and uncharged D-dimensional black hole is
related to its Schwarzschild radius $r_s$ through
\cite{5Dradius,BHTEV}:
\begin{equation}
M_{BH}=r_s^{D-3}
(\sqrt{\pi}M_D)^{D-3}M_D\frac{D-2}{8\Gamma(\frac{D-1}{2})},
\label{5drad}
\end{equation}
where the D-dimensional Planck mass $M_D$ is related to the
D-dimensional gravitational constant $G_D$ and the usual
4-dimensional Planck mass $M_p$ through:
\begin{eqnarray}
M_{D}=G_D^{-\frac{1}{D-2}}, \nonumber\\
M_p^2=M_D^{D-2}V_{D-4},\label{m5m4}
\end{eqnarray}
with $V_{D-4}$ the volume of the extra-dimensional space
\cite{5Dradius}.

If $\rld$ is the bulk vacuum energy, then application of
holographic dark energy in the bulk gives:
\begin{equation}
\rld {\text{Vol}}({\mathcal{S}}^{D-2})\leq r^{D-3}
(\sqrt{\pi}M_D)^{D-3}M_D\,\frac{D-2}{8\Gamma(\frac{D-1}{2})},
\label{resHDE}
\end{equation}
where ${\text{Vol}}({\mathcal{S}}^{D-2})$ is the volume of the
maximal hypersphere in a $D$-dimensional spacetime, given from:
\begin{equation}
{\text{Vol}}({\mathcal{S}}^{D-2})=A_D\,r^{D-1} \label{v2k},
\end{equation}
with
\begin{eqnarray}
A_D=\frac{\pi^{\frac{D-1}{2}}}{\left(\frac{D-1}{2}\right)!}\nonumber,\\
 A_D=\frac{\left(\frac{D-2}{2}\right)!}{\left(D-1\right)!} 2^{D-1}\, \pi^{\frac{D-2}{2}},
\label{ad}
\end{eqnarray}
for $D-1$ being even or odd respectively. Therefore, by saturating
inequality (\ref{resHDE}) introducing $L$ as a suitable large
distance (IR cut-off) and $c^2$ as a numerical factor, the
corresponding vacuum energy is,  as usual, viewed as holographic
dark energy:
\begin{equation}
\rld =c^2 (\sqrt{\pi}M_D)^{D-3}M_D A_D^{-1}
\frac{D-2}{8\Gamma(\frac{D-1}{2})}\,L^{-2} \label{resHD2}.
\end{equation}
As was mentioned in \cite{manos.restored}, the ``suitable large
distance'' which is used in the definition of $L$ in
(\ref{resHD2}) could be the Hubble radius \cite{interacting0},
proportional to the square root of the Hubble radius
\cite{Guberina}, the particle horizon \cite{Susskind.rh}, the
future event horizon \cite{Li,Hsu,Gong}, or the radius of the
event horizon measured on the sphere of the horizon \cite{Setare}
(see also \cite{Setare11} for the corresponding formulation in
Chaplygin gas and tachyon holographic models). For a flat Universe
the future event horizon is the most suitable ansatz and
furthermore it is the only one that fits holographic statistical
physics, namely the exclusion of those degrees of freedom of a
system that will never be observed by the effective field theory
\cite{Enqvist}.

\section{Holographic Dark Energy in General 5D Braneworld models with a Gauss-Bonnet Term in the Bulk}\label{HDEBrane}

We are interested in applying bulk holographic dark energy in
general 5D braneworld models with a Gauss-Bonnet term in the bulk.
We consider an action of the form \cite{Charmousis,TetradisGB}:
\begin{equation}
 S=\int d^4xdy\sqrt{-g}\left(M_5^3R-\rlf+M^3_5\alpha{\cal{L}}_{GB}\right)+\int
 d^4x\sqrt{-\gamma}\,\left({\cal{L}}_{br}^{mat}-V+r_cM^3_5R_4\right).
\label{action}
\end{equation}
In the first integral $M_5$ is the 5D Planck mass, $\rlf$ is the
bulk cosmological constant which is identified as the bulk
holographic dark energy, and $R$ is the curvature scalar of the 5D
bulk spacetime with metric $g_{AB}$. As usual,
\begin{equation}
 {\cal{L}}_{GB}=R^2-4R_{AB}R^{AB}+R_{ABCD}R^{ABCD}
\label{GBeq}
\end{equation}
is the Gauss-Bonnet term with coupling constant $\alpha$, and
$R_{ABCD}$, $R_{AB}$ are respectively the Riemann and Ricci
tensors. In the second integral $\gamma$ is the determinant of the
induced 4D metric $\gamma_{\alpha\beta}$ on the brane, $V$ is the
brane tension and ${\cal{L}}_{br}^{mat}$ is an arbitrary brane
matter content. Lastly, we have allowed for an induced gravity
term on the brane, arising from radiative corrections, with $r_c$
its characteristic length scale and $R_4$ the 4D curvature scalar
\cite{inducedgrav,Zamarias,Tetradis.ind.w1}.

In order to acquire the cosmological evolution on the brane we use
the Gaussian normal coordinates with the following metric form
\cite{tetradisbulk1,manos.mirage}:
\begin{equation}
ds^2=-m^2(\tau,y)d\tau^2+a^2(\tau,y)\,d\Omega^2_k+dy^2.
\label{metric}
\end{equation}
The brane is located at $y=0$, we impose a $Z_2$-symmetry around
it, $m(\tau,y=0)=1$ and $d\Omega^2_k$ stands for the metric in a
maximally symmetric 3-dimensional space with $k=-1,0,+1$
parametrizing its spacial curvature. Although we could assume a
general matter-field content \cite{manos.param}, we consider a
brane-Universe containing a perfect fluid with equation of state
$p=w\rho$. In this case, and after integration of the $00$ and
$ii$ components of the 5D Einstein equations around the brane, the
low-energy ($\rho\ll V$) brane cosmological evolution is governed
by the following equation \cite{TetradisGB,inducedgrav,Zamarias}
(see also \cite{sigkr.liseis} for similar brane solutions):
\begin{equation}
H^2+\frac{k}{a^2}=\left(72M^6_5-16\alpha\,\rlf
M_5^3+6r_cVM^3_5\right)^{-1}\,V\rho+\frac{V^2}{144M^6_5}-\frac{1-\sqrt{1+\tilde{\Lambda}}}{36\alpha}\,\left(2+\sqrt{1+\tilde{\Lambda}}\right)^2,
\label{fried1}
\end{equation}
where
\begin{equation}
\tilde{\Lambda}=2\alpha\rlf/3M_5^3. \label{Ltild}
\end{equation}
In (\ref{fried1}) $a$ stands as usual for the brane scale factor.
In order to acquire a form consistent with conventional 4D
Friedmann equation we make the identification:
\begin{equation}
V=\frac{72M_5^3}{\frac{3}{8\pi}\frac{M_p^2}{M_5^3}-6r_c},
\label{Vm}
\end{equation}
and we define
\begin{equation}
V_1(\alpha,\rlf)=\frac{2\alpha\rlf\left(\frac{3}{8\pi}\frac{M_p^2}{M_5^3}-6r_c\right)}
{9M_5^6\left(\frac{3}{8\pi}\frac{M_p^2}{M_5^3}\right)^2-2\alpha\rlf\frac{3}{8\pi}M_p^2\left(\frac{3}{8\pi}\frac{M_p^2}{M_5^3}-6r_c\right)},
 \label{V1m}
\end{equation}
where $M_p$ is the 4D Planck mass. In this case brane evolution
equation (\ref{fried1}) becomes:
\begin{equation}
H^2+\frac{k}{a^2}=\frac{8\pi}{3M_p^2}\,\rho+V_1(\alpha,\rlf)\,\rho+\frac{8\pi}{3M_p^2}\,\rl,\label{fried2}
\end{equation}
where the (effective in this higher-dimensional model) 4D dark
energy is:
\begin{equation}
\rl\equiv\rho_{\Lambda4}=\frac{3M_p^2}{2\pi(\frac{1}{8\pi}\frac{M_p^2}{M^3_5}-2r_c)^2}-
\frac{M_p^2}{96\pi
\alpha}\left(1-\sqrt{1+\tilde{\Lambda}}\right)\left(2+\sqrt{1+\tilde{\Lambda}}\right)^2.
\label{rl4}
\end{equation}
In the equations above $\rlf$ is the 5D bulk holographic dark
energy, which according to (\ref{resHD2}) is given by:
\begin{equation}
\rlf=c^2\frac{3}{4\pi}M_5^3L^{-2}\label{rlf}.
\end{equation}
Relations (\ref{Ltild})-(\ref{rlf}) describe the low-energy
(late-time) cosmological evolution on the brane. Similarly to
\cite{manos.restored,manos.movingbranes} the holographic nature of
$\rlf$ is the cause of the holographic nature of $\rl$. Finally,
the 5D Planck mass $M_5$ is related to the standard 4D $M_p$
through $M_5^3=M_p^2/L_5$ (according to (\ref{m5m4})), with $L_5$
the volume (size) of the extra dimension.

Let us make some comments here. The above expressions in the limit
$\alpha\rightarrow0$ (where $\tilde{\Lambda}\rightarrow0$ and
$V_1(\alpha,\rlf)\rightarrow0$) tend smoothly to those analyzed in
\cite{manos.restored}. However, in the presence of the
Gauss-Bonnet term ($\alpha\neq0$) we observe an interesting
interacting behavior. Indeed, in (\ref{fried2}) there is a
coupling between $\rho$ and $V_1(\alpha,\rlf)$, that is a term
depending on $\rlf$ and therefore on $\rl$. We mention that the
coupling between $\rho$ and $\rl$ arises naturally through the
full 5D dynamics and the use of bulk holographic dark energy, and
it is not a result of an arbitrary introduction by hand, which is
the usual case in interacting holographic dark energy in the
literature \cite{interacting0,interacting} even in the case where
a Gauss-Bonnet term is present \cite{interactingGB}.

Our final goal is to find the relation between $\rl$ and the
metric scale factor $a$ of the brane. However, the complex form of
the above equations makes it impossible to acquire such an
expression analytically. Therefore, in the following we describe
the necessary approximations. Firstly, as we have already
mentioned, according to (\ref{m5m4}) $M_5^3=M_p^2/L_5$ with $L_5$
the volume of the extra dimension. In this work we assume that
$L_5$ is arbitrary large (but not infinite), i.e. it is larger
than any other length of the model, thus leaving brane evolution
unaffected by the bulk size or bulk boundaries and this is the
reason for the single-brane consideration. Therefore, in the
calculations below we impose $M_p^2/M_5^3=L_5\gg r_c$ and
$1/L_5\rightarrow0$. The role of the bulk size was investigated in
\cite{manos.movingbranes}. Secondly, we expand (\ref{V1m}) and
(\ref{rl4}) in terms of the Gauss-Bonnet coupling $\alpha$ and we
keep only the linear term. Actually this is also a consistency
requirement since, in heterotic string theory background, the
Gauss-Bonnet form is the leading order quantum correction to
gravity, i.e we have already kept only linear terms in $\alpha$
\cite{green}. These steps lead to:
\begin{equation}
V_1(\alpha,\rlf)\equiv
V_1(\alpha,L)=\frac{4}{9}\frac{c^2}{M_p^2}\,\alpha L^{-2}+{\cal
O}(\alpha^2),
 \label{V1m2}
\end{equation}
\begin{equation}
\rl=3c^2\frac{1}{128\pi^2}M_p^2\,L^{-2}\left(1+\alpha\,\frac{c^2}{24\pi}\,L^{-2}\right)+{\cal
O}(\alpha^2). \label{rl4b1}
\end{equation}
Finally, we have to determine the cosmological length $L$ which is
present in the bulk holographic dark energy expression (\ref{rlf})
and has been transferred to relations (\ref{V1m2}),(\ref{rl4b1}),
too. In the following we will consider a flat Universe, in order
to safely use the future event horizon to define $L$, without
entering into the relevant discussion of the literature concerning
the IR cut-off in non-flat cases
\cite{Li,Guberina,Setare,Hsu,Gong}. However, the model of the
present work, such as the majority of braneworld models of the
literature, is not maximally isotropic and this feature makes the
analytical calculation of the future event horizon an impossible
task. In this anisotropic case we can alteratively use the 4D
future event horizon $R_h$ (the 4D spacetime is the maximally
isotropic subspace of the model), without losing the qualitative
behavior of the observables. Fortunately, the calculations in the
simple case without a Gauss-Bonnet term \cite{manos.restored},
showed that the use of the 4D future event horizon leads to
identical quantitative results comparing to those obtained within
the traditional holographic dark energy
\cite{Li,hol1,Gong,Guberina,Setare,Setare11}.

Using the above approximations we obtain the following form for
the effective 4D holographic dark energy:
\begin{equation}
\rl=3c^2\frac{1}{128\pi^2}M_p^2\,R_h^{-2}\left(1+\alpha\,\frac{c^2}{24\pi}\,R_h^{-2}\right),
\label{rl4b}
\end{equation}
and substitution to Friedmann equation (\ref{fried2}), for the
flat-Universe case, gives:
\begin{equation}
H^2=\frac{8\pi}{3M_p^2}\,\rho\left(1+\alpha\,\frac{c^2}{6\pi}R_h^{-2}\right)+
\frac{c^2}{16\pi}R_h^{-2}\left(1+\alpha\,\frac{c^2}{24\pi}R_h^{-2}\right).
 \label{fried3}
\end{equation}
In these relations, the 4D future event horizon $R_h$ is given as
usual by:
\begin{equation}
R_h=a\int_{a}^{\infty}\frac{da'}{Ha'^2}. \label{Rh}
\end{equation}
Finally, we have to insert in (\ref{fried3}) the known form for
$\rho(a)$, namely $\rho=\rho_0a^{-3}$, with $\rho_0$ its present
value.

The aforementioned integral equations determine completely the
brane evolution, in the low energy limit, and up to first order in
terms of the Gauss-Bonnet coupling $\alpha$. In the limit
$\alpha\rightarrow0$ these expressions coincide with those
extracted in \cite{manos.restored}. However, in the presence of
the Gauss-Bonnet term the implications are significant. Firstly,
4D holographic dark energy $\rl$, apart from the usual squared
holographic term, acquires a quartic correction. Secondly, matter
density $\rho$ is coupled with a holographic term $\propto
R_h^{-2}$, which is a result of $\rho$-$\rl$ interaction of
equation (\ref{fried2}).

Analytical solution of equations (\ref{rl4b})-(\ref{Rh}), namely
finding $H(a)$, then $R_h(a)$, and finally $\rl(a)$, is
impossible. However, we are not interested in investigating the
complete evolution but only in revealing the form of $\rl(a)$.
Thus, we generalize Li's steps to construct a differential
equation using $\Omega_\Lambda$ as the unknown function \cite{Li}.

Firstly, we insert the usual variables:
$\oll=\frac{8\pi\rl}{3M_p^2 H^2}$, $\omm=\frac{8\pi\rho}{3M_p^2
H^2}$. Relation (\ref{rl4b}) then gives:
\begin{equation}
R_h=\frac{c_1}{\sqrt{\oll}H}+\alpha \,c_2\sqrt{\oll}H
 \label{rh2}
\end{equation}
up to ${\cal O}(\alpha^2)$, with $c_1=\frac{c}{4 \sqrt{\pi}}$ and
$c_2=\frac{c}{12\sqrt{\pi}}$. Inserting this form in (\ref{Rh})
and using the variable $x=\ln\alpha$ we obtain:
\begin{equation}
\int^\infty_x\frac{dx}{Ha}=\frac{1}{a}\,\left(\frac{c_1}{\sqrt{\oll}H}+\alpha
\,c_2\sqrt{\oll}H\right).
 \label{rh3}
\end{equation}
Similarly, using $\oll$, $\omm$, and $R_h$ from (\ref{rh2}),
Friedmann equation (\ref{fried2}) (with $V_1(\alpha,\rlf)$ given
by (\ref{V1m2})) up to ${\cal O}(\alpha^2)$ writes:
\begin{equation}
1-\oll=\omm\left(1+\alpha\,2c_3\oll H^2\right),
 \label{fried4}
\end{equation}
where $c_3=32\pi/3$. In order to proceed forward we have to assume
an explicit $\omm(a)$ dependence. In the interacting case at hand
this should be different from the known $\sim a^{-3}$ behavior of
standard cosmology. However, in our model the $\rho$-$\rl$
interaction is downgraded by the extra-dimensional size as can be
seen in (\ref{V1m}) or equivalently in (\ref{V1m2}). Therefore,
the deviation from conventional evolution will not be significant
and we can use $\omm=\omm^0 H_0^2H^{-2}a^{-3}$ with $\omm^0$ and
$H_0$ the present values. Thus, we obtain:
\begin{equation}
\frac{1}{Ha}=\frac{\sqrt{a}\sqrt{1-\oll}}{\sqrt{\omm^0}H_0}\left[1-\alpha\,c_3\oll
\frac{\omm^0H_0^2}{a^3(1-\oll)}\right].
 \label{help}
\end{equation}
Finally, substituting this relation to (\ref{rh3}) and taking
derivative with respect to $x$, up to ${\cal O}(\alpha^2)$ we
acquire the following differential equation:
\begin{equation}
\oll'=Q_1(\oll)+\alpha \,Q_2(\oll,a),
 \label{diff}
\end{equation}
where
\begin{equation}
Q_1(\oll)=\oll^2(1-\oll)\left[\frac{1}{\oll}+\frac{2}{c_1\sqrt{\oll}}\right],
 \label{q1}
\end{equation}
and
\begin{equation}
Q_2(\oll,a)=\frac{\omm^0H_0^2}{c_1a^3}\left\{\left(c_2-c_3c_1\right)
\left[-5\oll^2+Q_1(\oll)\left(\frac{1}{\oll}-1\right)^{-1}\right]-2c_3\oll^{5/2}\right\},
 \label{q2}
\end{equation}
and the prime denotes the derivative with respect to $x$. Note
that in the limit $\alpha\rightarrow0$, differential equation
(\ref{diff}) tends smoothly to that obtain by Li in \cite{Li},
namely $\oll'=Q_1(\oll)$, and can be easily solved analytically.
In the $\alpha\neq0$ case of the present work such an exact
solution is impossible. However, under the identification
$\rl(a)\sim a^{-3(1+w_\Lambda)}$, we can extract the form of
$w_\Lambda(z)$ at late times, i.e. at small $z$, with
$z=\frac{a_0}{a}-1$ and $a_0$ the value of $a$ at present time
(for simplicity we set $a_0=1$). We proceed as follows:

Firstly, expanding $\ln\rl$ we obtain:
\begin{equation}
\ln\rl=\ln\rl|_0+\frac{d\ln\rl}{d\ln a}|_0\,\ln a
+\frac{1}{2}\frac{d^2\ln\rl}{d(\ln a)^2}|_0\,(\ln a)^2+{\cal
O}\left((\ln a)^3\right),
 \label{expand}
\end{equation}
where the derivatives are calculated at the present time $a_0=1$
\cite{Li}. Therefore, through $\rl(a)\sim a^{-3(1+w_\Lambda)}$ we
make the identification:
\begin{equation}
w_\Lambda=-1-\frac{1}{3}\left[\frac{d\ln\rl}{d\ln
a}|_0+\frac{1}{2}\frac{d^2\ln\rl}{d(\ln a)^2}|_0\,\ln a+{\cal
O}\left((\ln a)^2\right)\right].
 \label{wl}
\end{equation}

 Now, using Friedmann equation (\ref{fried4}), and the
 expressions $\oll=\frac{8\pi\rl}{3M_p^2 H^2}$ and $\omm=\omm^0 H_0^2H^{-2}a^{-3}$, we find:
\begin{equation}
\rl=\frac{3M_p^2\oll}{8\pi}\frac{\omm^0H_0^2}{a^3(1-\oll)}\left[1+\alpha\,
2c_3\oll\frac{\omm^0H_0^2}{a^3(1-\oll)} \right],
 \label{rll}
\end{equation}
up to ${\cal O}(\alpha^2)$. Therefore, differentiating this
relation with respect to $\ln a =x$, and using (\ref{diff}) for
the calculation of the derivatives, we finally obtain the
following $w_\Lambda$ expression:
\begin{equation}
w_\Lambda(z)=w_0+ w_1 z+\alpha(w_2+ w_3z),
 \label{wl2}
\end{equation}
where
\begin{equation}
w_0=-\frac{1}{3}-\frac{2}{3c_1}\sqrt{\Omega_\Lambda^0},
 \label{w0}
\end{equation}
\begin{equation}
w_1=\frac{1}{6c_1}\sqrt{\Omega_\Lambda^0}(1-\Omega_\Lambda^0)
\left(1+\frac{2\sqrt{\Omega_\Lambda^0}}{c_1}\right),
 \label{w1}
\end{equation}
\begin{equation}
w_2=\frac{2}{3c_1}\frac{\Omega_\Lambda^0}{1-\Omega_\Lambda^0}
\left[b_1c_1+2b_2b_3c_1-\sqrt{\Omega_\Lambda^0}(b_1+
b_2b_3-c_1c_3b_2)\right],
 \label{w2}
\end{equation}
\begin{eqnarray}
w_3=-\frac{1}{6c_1^2}\frac{\Omega_\Lambda^0}{1-\Omega_\Lambda^0}
\left\{-4(b_1+2b_2b_3)c_1^2+c_1(7b_1+15b_2b_3-3b_2c_1c_3)\sqrt{\Omega_\Lambda^0}+
  \right.\nonumber\\+\left.
(8b_2c_1c_3-6b_1-8b_2b_3)\Omega_\Lambda^0+
c_1[b_1-b_2(3b_3+c_1c_3)](\Omega_\Lambda^0)^{3/2}+\right.\nonumber\\+\left.
2[b_1+2b_2(b_3-c_1c_3)](\Omega_\Lambda^0)^2 \right\}.
 \label{w3}
\end{eqnarray}
In the expressions above we have used the constants $b_1=2c_3c_4$,
$b_2=c_4/c_1$ and $b_3=c_2-c_3c_1$, where $c_4=\omm^0H_0^2$.
Moreover, since $a_0=1$, we have replaced $\ln a=-\ln(1+z)\approx
-z$. Finally, $\Omega_\Lambda^0$ is the present value of
$\Omega_\Lambda$.

Relation (\ref{wl2}) is the main result of this work and provides
the Gauss-Bonnet correction to the corresponding result of
\cite{manos.restored}. Both investigations are formulated in the
framework of bulk holographic dark energy. Therefore, although in
the limit $\alpha\rightarrow0$, (\ref{wl2}) coincides with Li's
expression in \cite{Li}, namely $w_\Lambda(z)=w_0+ w_1 z$, the
physical explanation in the present case comes through the 5D
holographic consideration. This is the reason of the difference in
constants between this work and \cite{Li}.

From (\ref{wl2}) it becomes obvious, that according to the value
of  $c$ which is present in $\rlf$-relation (\ref{rlf}), of $c_4$
and of the Gauss-Bonnet coupling constant $\alpha$, one can obtain
a 4D holographic dark energy behaving as phantom \cite{phantom},
quintessence or quintom \cite{quintom0,quintom}, i.e crossing the
phantom divide $w_\Lambda=-1$ \cite{quintoma,Tetradis.ind.w1}
during the evolution. Additionally, one can use observational
results concerning dark energy evolution
\cite{observHDE1,observHDEaaa} in order to estimate the bounds of
the constant $c$ of \cite{Li}, i.e the bounds of $c_1$ of the
present work. In particular, observational data from type Ia
supernovae give the best-fit value $c_1=0.21$  within 1-$\sigma$
error range \cite{observHDEIa}, while those from the X-ray gas
mass fraction of galaxy clusters lead to $c_1=0.61$ within
1-$\sigma$ \cite{observHDExray}. Similarly, combining data from
type Ia supernovae, cosmic microwave background radiation and
large scale structure give the best-fit value $c_1=0.91$ within
1-$\sigma$ \cite{observHDECMB}, while combining data from type Ia
supernovae, X-ray gas and baryon acoustic oscillation lead to
$c_1=0.73$ as a best-fit value within 1-$\sigma$ \cite{observHDE}.
Inserting this range of $c_1$ values into our model one finds that
$w_0<-1$ and $w_1>0$, thus, within 1-$\sigma$, he obtains a
quintom-type holographic dark energy. Furthermore, $w_2<0$ while
$w_3>0$ and therefore the Gauss-Bonnet contribution decreases the
present value of $w_\Lambda$, while it increases the growing rate
of $w_\Lambda(z)$ with $z$, in comparison with the case where such
a term is absent. However, the quantitative correction of the
$\alpha\neq0$ case will be very small, for reasonable $c_4$
values. The reason is that, as we have mentioned, the $\rho$-$\rl$
coupling, which arose naturally as a term $V_1(\alpha,\rlf)\,\rho$
in (\ref{fried2}), is downgraded by the extra-dimensional size as
can be seen in (\ref{V1m}) or equivalently in (\ref{V1m2}) (where
we acquire a $L^2$ in the denominator). Thus, making the
assumption that $L_5$ is arbitrary large we downgrade the
Gauss-Bonnet correction, too. It should be interesting to
investigate the case where the bulk-size is smaller than the
future event horizon, as in the two-brane model of
\cite{manos.movingbranes,twobrane}, but with the inclusion of a
Gauss-Bonnet term. The subject is under investigation. Finally,
note that the role of the Gauss-Bonnet term on the $w=-1$ crossing
has been investigated both in conventional 4D \cite{GB4D.w1} and
in braneworld frameworks \cite{GBbulk.w1,Zamarias}. The novel
feature of our work is the combined investigation of such a term
with the bulk holographic dark energy.

\section{Discussion-Conclusions}\label{discussion}

In this work we apply bulk holographic dark energy in a general
braneworld model, with an induced gravity term and a perfect fluid
on the brane, and a Gauss-Bonnet term in the bulk. Such a
generalized bulk version of holographic dark energy is necessary
if we desire to match the successes of brane cosmology in both
theoretical and phenomenological-observational level, with the
successful, simple, and inspired by first principles, notion of
holographic dark energy in conventional 4D cosmology. In
particular, as we showed in \cite{manos.restored}, the bulk space
is the natural framework for the cosmological application,
concerning dark energy, of holographic principle, since it is the
maximally-dimensional subspace that determines the properties of
quantum-field and gravitational theory, and the black hole
formation. Subsequently, this bulk holographic dark energy will
give rise to an effective 4D dark energy with ``inherited"
holographic nature, and this one will be present in the effective
Friedmann equation.

Taking the Gauss-Bonnet combination into account, a $\rho$-$\rl$
coupling appears in the Friedmann equation of the brane. We
mention that this term arises naturally and is not a result of an
inclusion by hand, which is the usual case of 4D interacting
holographic dark energy in the literature
\cite{interacting0,interacting,interactingGB}. This fact makes
bulk holographic dark energy in the Gauss-Bonnet framework an
interesting subject for further investigation.

Examining the low-energy (late-time) evolution of the
aforementioned model, we acquire the relation of $w_\Lambda(z)$ up
to ${\cal O}(\alpha^2)$ and ${\cal O}(z^2)$. In the limit
$\alpha\rightarrow0$ we re-obtain the results of
\cite{manos.restored} and those of conventional 4D calculations
\cite{Li,hol1,Gong,Guberina,Setare,Setare11}, although in the 5D
study the interpretation and explanation of these results  is
fundamentally different. In the presence of Gauss-Bonnet
combination, and taking into account the constraints on the values
of the constants by observational data, we find that the effective
4D holographic dark energy behaves as a quintom, i.e it crosses
the phantom divide $w_\Lambda=-1$ during the evolution. In
particular, we observe that the presence of a non-zero $\alpha$
makes the current value of $w_\Lambda$ smaller, while it increases
its growing rate with $z$, comparing to the $\alpha=0$ case.
However, the corresponding quantitative correction is very small
due to the diminution of the $\rho$-$\rl$ coupling by the
arbitrary large extra-dimensional size. Yet, it should be
interesting to investigate the case where the bulk size is smaller
than the future event horizon. Then, the $\rho$-$\rl$ coupling
would be significant and we would naturally acquire the advantages
of interacting holographic dark energy, such as the coincidence
problem solution, and the corresponding effects on $w_\Lambda(z)$.
\\

\paragraph*{{\bf{Acknowledgements:}}}
The author is grateful to  G.~Kofinas, K.~Tamvakis, N. Tetradis,
F. Belgiorno, B. Brown, S. Cacciatori, M. Cadoni, R.~Casadio,
G.~Felder, A.~Frolov, B. Harms, N.~Mohammedi, M.~Setare and
Y.~Shtanov for useful discussions.

\end{document}